# Simulation of the Efficiency of CdS/CdTe Tandem Multi-Junction Solar Cells

Ashrafalsadat S. Mirkamali[1,2], Khikmat Kh. Muminov[1]

[1]S.U.Umarov Physical-Technical Institute, Academy of Sciences of the Republic of Tajikistan, 299/1 Aini Ave, Dushanbe 734062, Tajikistan
e-mail: khikmat@inbox.ru
[2]Permanent address: Department of Science and Engineering, Behshahr Branch, Islamic Azad University, Behshahr, Iran
e-mail: ash.mirkamali@gmail.com

**Abstarct**

In this paper we study CdS/CdTe solar cells by means of AMPS-1D software. First we study the effect of thickness of semiconductor layers on the output parameters of the CdS/CdTe solar cell, such as density of short-circuit current, open circuit voltage, fill factor and efficiency. Numerical simulation shows that the highest efficiency of single-junction CdS/CdTe solar cell equal to 18.3% is achieved when the CdTe layer thickness is 1000 nm and a CdS layer is 60 nm. Then, in order to obtain the maximal value of the efficiency, new tandem multi-junction structure consisting of layers of two solar cells connected with each other back to back are designed and engineered taking into account the results obtained for the single-junction solar cells. Numerical simulations show that its highest efficiency in 31.8% can be obtained when the thickness of CdS p-layer is equal to 50 nm, and the thickness of the CdS n-layer is equal to 200 nm, while thicknesses of the CdTe n-layer and CdTe p-layer are kept fixed and equal to 3000 nm and 1000 nm, respectively.

## INTRODUCTION

It is well known theoretically, that for the cadmium telluride CdTe solar cells, the minimal thickness of the CdTe film, required to absorb 99% of incident photons with energies greater than the band gap $E_g$, is about 1-2 microns [1, 2]. Reducing the thickness of the CdTe absorber layer is reasonable not only in terms of cost reduction of material in the manufacturing process, but also can improve the efficiency of the solar cell due to reduction of recombination losses by volume [3, 4].

However, although the thin films are more compact and free of micropores, the growth control and even repeated crystallization should be provided for their preparation [4]. So, in 1982, Tyan et al published an interesting paper on the CdTe/CdS thin-film solar cells having the efficiency of about 10% [5]. Subsequently, the efficiency of 15.8% was achieved by Ferekides et al. [6]. Finally, a group of the US researchers in the National Renewable Energy Laboratory have reported a record efficiency of 16.5% [7]. This record-breaking solar CdS/CdTe cell with the efficiency of 16.5% used a modified structure of the CTO/ZTO/CdS/CdTe with the CdS layer with the thickness of 0.1 microns and the CdTe layer of 10 microns. It was produced using three different techniques: close space sublimation (CSS) for deposition of CdTe film, the method of chemical bath deposition (CBD) to form a CdS film and magnetron sputtering for the remaining layers. The resulting record efficiency solar cells (16.5%) is slightly more than half of the theoretical limit of 29%, but it is believed that practical CdTe devices with an efficiency of 18-19% can be achieved in the near future [8]. To reduce the absorption and to minimize the surface recombination current in the CdS/CdTe solar cells, a reduction of CdS layer thickness has been consistently carried out to 1000 Å and less [9]. At this thickness, they have a low performance due to increasing the likelihood of shunting.

Malaysian group [10] in 2011 has developed a $Zn_xCd_{1-x}S/CdTe$ solar cell. In order to study the effect of thin-film CdTe absorber a numerical simulation by use AMPS-1D software has been carried out. The thickness of the CdTe absorber layer varied in the range from 100 nm to 6000 nm at this simulation. Further reduction in the thickness of CdTe layer from 1 micron showed that the short circuit current density decreases slowly and the open circuit voltage remains practically unchanged, while the fill factor increases until a 600 nm [10]. Results showed that the efficiency of 19.5% could be achieved by use of economical ultrathin $Zn_xCd_{1-x}S/CdTe$ solar cell structures, where the CdTe layer has the thickness of 1 micron, and the thickness of $Zn_xCd_{1-x}S$ is 60 nm, with the ZnO or $Zn_2SO_4$ buffer layers of 100 nm thickness. This solar cell has been studied in terms of stability to variation of temperature and also other properties of used materials have been investigated in the paper [11].

The purpose of this study is numerical simulation of CdS/CdTe in hetero-junction and multi-junction tandem solar cells and determining the optimal structure of the last, which would have a highest efficiency of conversion.

**THIN FILM CdS/CdTe SOLAR CELLS**

Thin film solar cells require less semiconductor material and it is easier to manufacture and assemble, they has less weight, they are flexible and less expensive than wafer solar cells. To date, there are a number of thin-film materials such as cadmium telluride CdTe, copper indium diselenide CIS, which are developed as materials for hetero-junction solar cells. Materials for production of thin-film heterojunction solar cells include such materials as gallium arsenide GaAs, indium phosphide InP, amorphous silicon a-Si, polycrystalline silicon, and so on. Such a material like CdTe is well deposited and well suited for large-scale production, however, cadmium is toxic. Small variations of CIS, obtainable by adding gallium gives a copper indium/gallium diselenide (CIGS), which shows the highest efficiency.

Cadmium telluride (CdTe) is a typical p-type semiconductor of II-VI group having an optical band gap of 1.5 eV, which is optimal one for photogeneration. The ability to uniform sputtering and a wide range of thermodynamic stability implies that the conventional sputtering process can be easily used for the deposition of the CdTe film on substrates with a large area [12].

Typical structure of the CdTe/CdTe solar cell is shown in Fig. 1. Its manufacturing usually begins on a glass substrate on which a thin layer of $In_2O_3$-$SnO_2$ (ITO) as a transparent conductive oxide (TCO) is deposited. The thickness of this layer is chosen such as to ensure both the conductivity and the optical transparency. As a window lyer there Cd n-type layer serve [12]. Then, on the layer enriched by tellurium various metals (Zn [13] Sb [14], Au [15], usually copper or Cu [15, 16]) are deposited with the graphite paste, so as to form a contact metal/metal telluride.

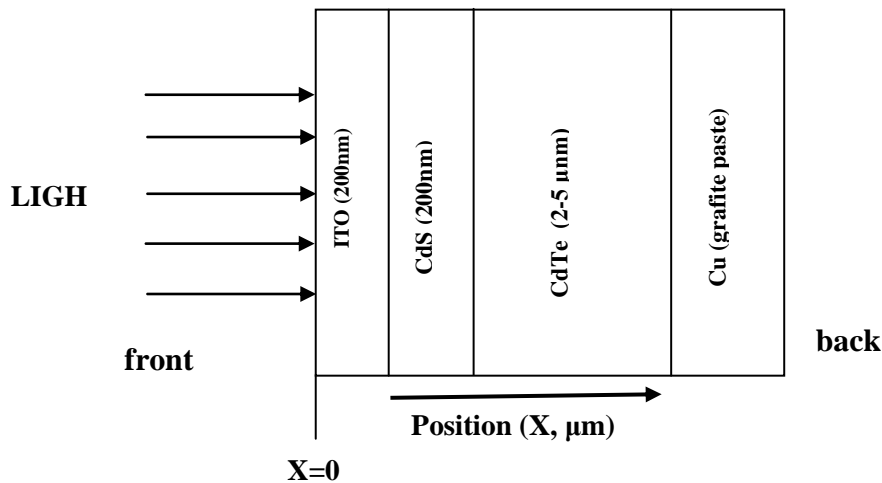

Figure 1. Typical structure of the CdTe/CdTe solar cell [13]

Numerical simulation of the solar cells is an important step in the analysis of their structure and properties, especially, in order to determine the influence of variations in the structure and properties of their individual parts on the efficiency of the solar cell as a whole. In general, such an analysis can be done by owning, sufficiently information about the characteristics of the input data. Optoelectronic properties of the materials from which the layers are produced, used in the numerical modeling, were taken from the paper of Gloeckler [18] devoted to thin-film photovoltaic cells. Profile of absorption (absorbance) was calculated by the Brownson et al. [19]. The values of materials parameters used in this simulation are shown in Table 1. In this study, as the material from which the front absorbent layer is made, i.e. the absorber material, is CdS with the bandgap (2.4 eV) to absorb ultra-violet region of the standard AM 1.5 solar spectrum. A layer of p-type in this study is the CdTe (1.5 eV). Transparent thin layer of indium oxide $In_2O_3$-$SnO_2$ with a bandgap of 3.60 eV (ITO) is used as the transparent conductive contacts, i.e. a material that covers the front surface of the oxide layer.

Table 1. The material parameters used in the numerical simulation of unijunction CdS / CdTe solar cell

| Material | Band gap (eV) | Conductivity type | Conduction Band | Valence Band | Electron Affinity (eV) | Electron Mobility (cm2 /v/s) | Hole Mobility (cm2 /v/s) | Free Carrier Concentration (cm$^{-3}$) | Relative Permittivity |
|---|---|---|---|---|---|---|---|---|---|
| ITO | 3.60 | N | 2.0*1020 | 1.8*1019 | 4.10 | 50.0 | 70.0 | 1.0*1020 | 2.0 |
| CdS | 2.40 | N | 2.2*1018 | 1.8*1019 | 4.0 | 25.0 | 100.0 | 1.1*1018 | 10.0 |
| CdTe | 1.50 | P | 8.0*1017 | 1.8*1019 | 3.90 | 40.0 | 320.0 | 2.0*1014 | 9.4 |

Theoretically, the minimum thickness of the film CdTe, required to absorb 99% of incident photons with energies greater than the band gap is approximately 1-2 mm [20, 21]. In this numerical simulation, the task of saving the required material and thus reduce the cost of solar cells based on CdS/CdTe by finding the optimum thickness of CdS and CdTe layers. To achieve these objectives, in a first step, the thickness of the CdTe ranged from 10 nm to 6000 nm, in order to find the thin absorber layer. The results of this simulation are shown in Fig.2. As ane can see from Fig.2, all the output parameters of the solar cell are approximately constant near 2000 nm for CdTe. However, when the thickness of CdTe is less than 1 micron all the output parameters of the solar cell is reduced, which is in good agreement with the results of other authors [22]. We have obtained a solar cell efficiency of 14.6% at a layer thickness of 1 micron CdTe. These results are in good agreement with the corresponding results published by other authors for CdTe solar cells [23].

Figure 2. The output characteristics of solar cells (short-circuit current density Jsc, open-circuit voltage Voc, fill factor FF and efficiency EFF) depending on the thickness of CdTe

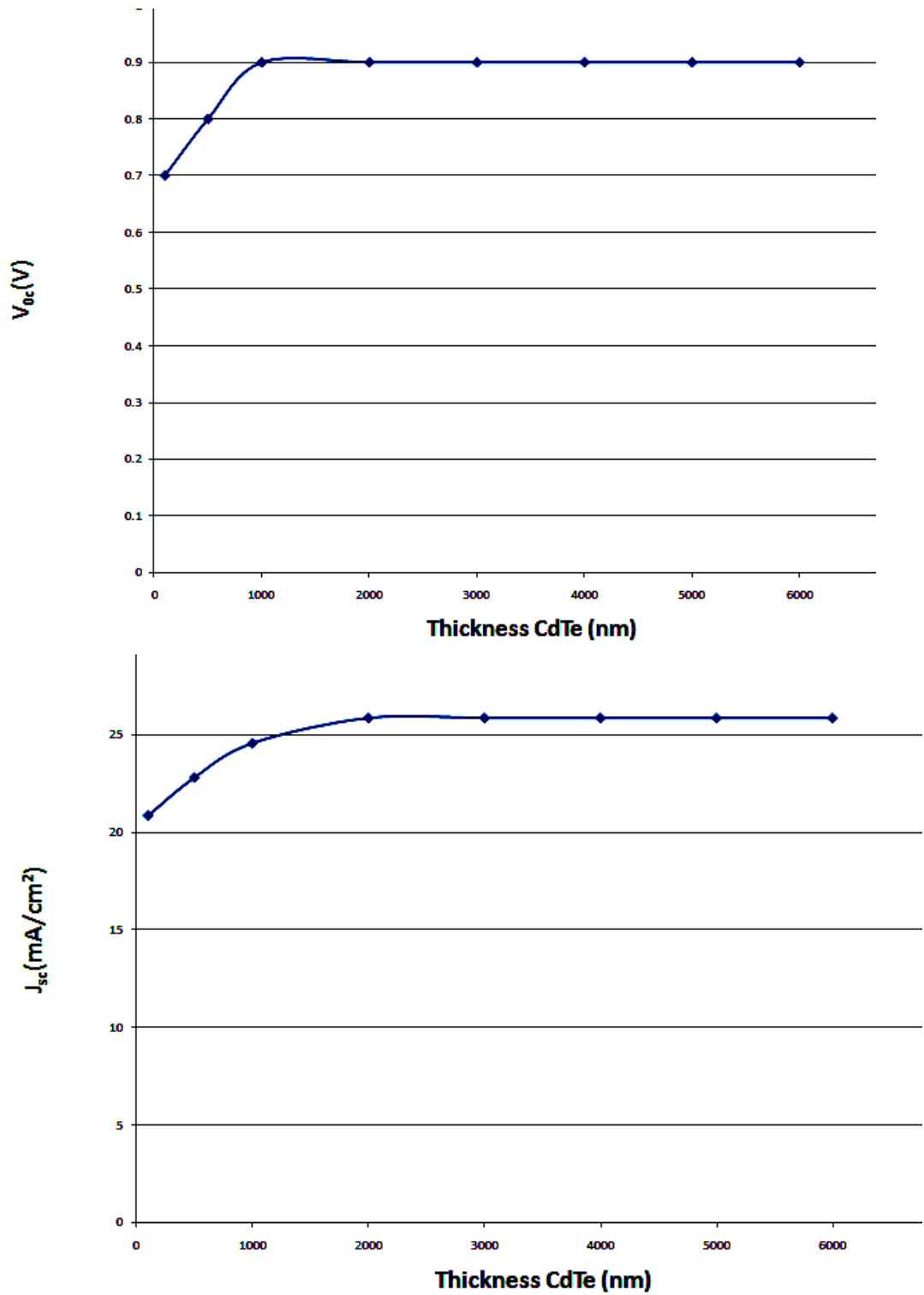

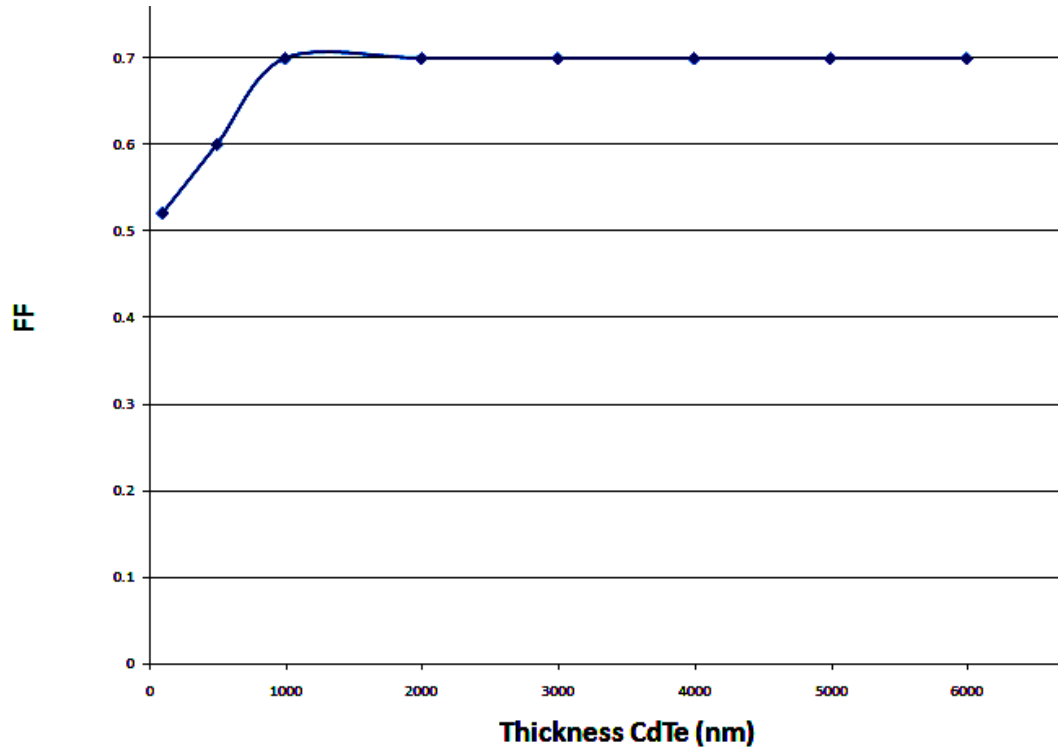

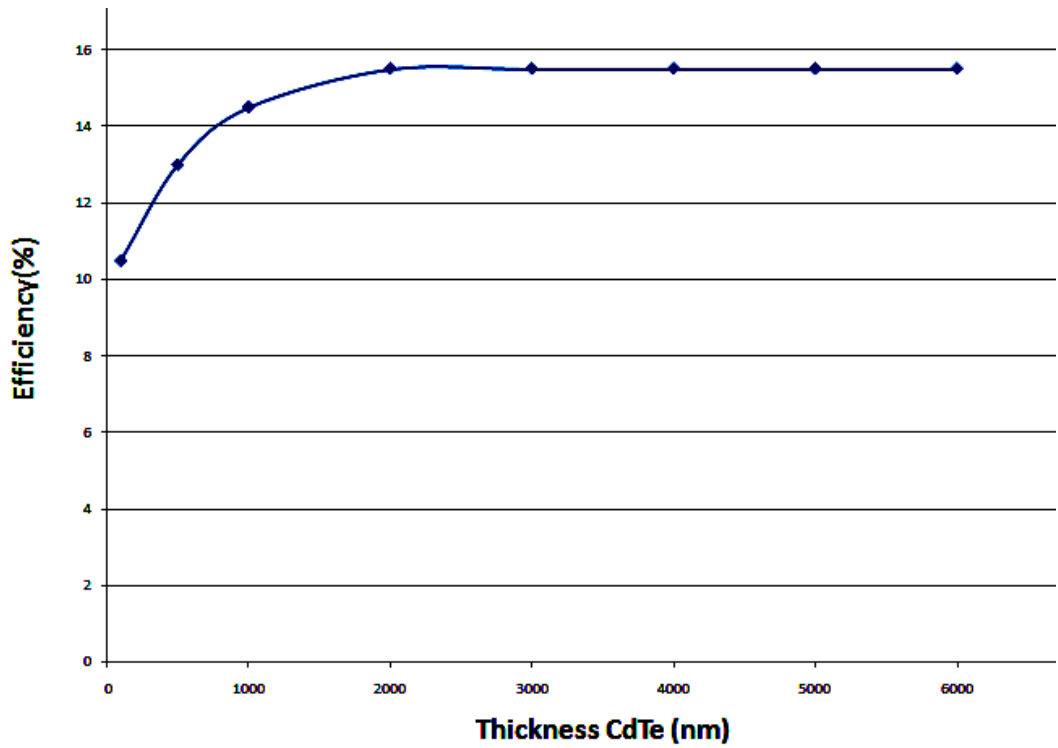

In the next step we carry out simulation of the CdTe solar cell based under other conditions, we put the CdTe layer thickness fixed and equal to 1000 nm, the CdS layer thickness ranges from 20 nm to 200 nm. In the Fig. 3 we show the effect of varying the thickness of CdS layer on the output parameters of the solar cell, such as the Jsc, Voc, FF and η, obtained by simulation on AMPS-1D. Note that these diagrams are obtained on the assumption that the optimum thickness of the CdS layer is 60 nm.

Thus, by means of numerical simulation, we determined that for the CdS/CdTe solar cells the highest efficiency equal to 18.3% is achieved when the thickness of the CdTe layer is 1000nm and thickness of CdS layer is 60 nm. This result is in good agreement with the experimental data obtained by Chou et al., [17], in which the maximum efficiency of the CdS/CdTe solar cells was 17.3%.

Fig. 3. Influence of the thickness of CdS on the output parameters (short-circuit current density Jsc, the open circuit voltage Voc, the fill factor FF and efficiency EFF) of the single-junction CdTe solar cell

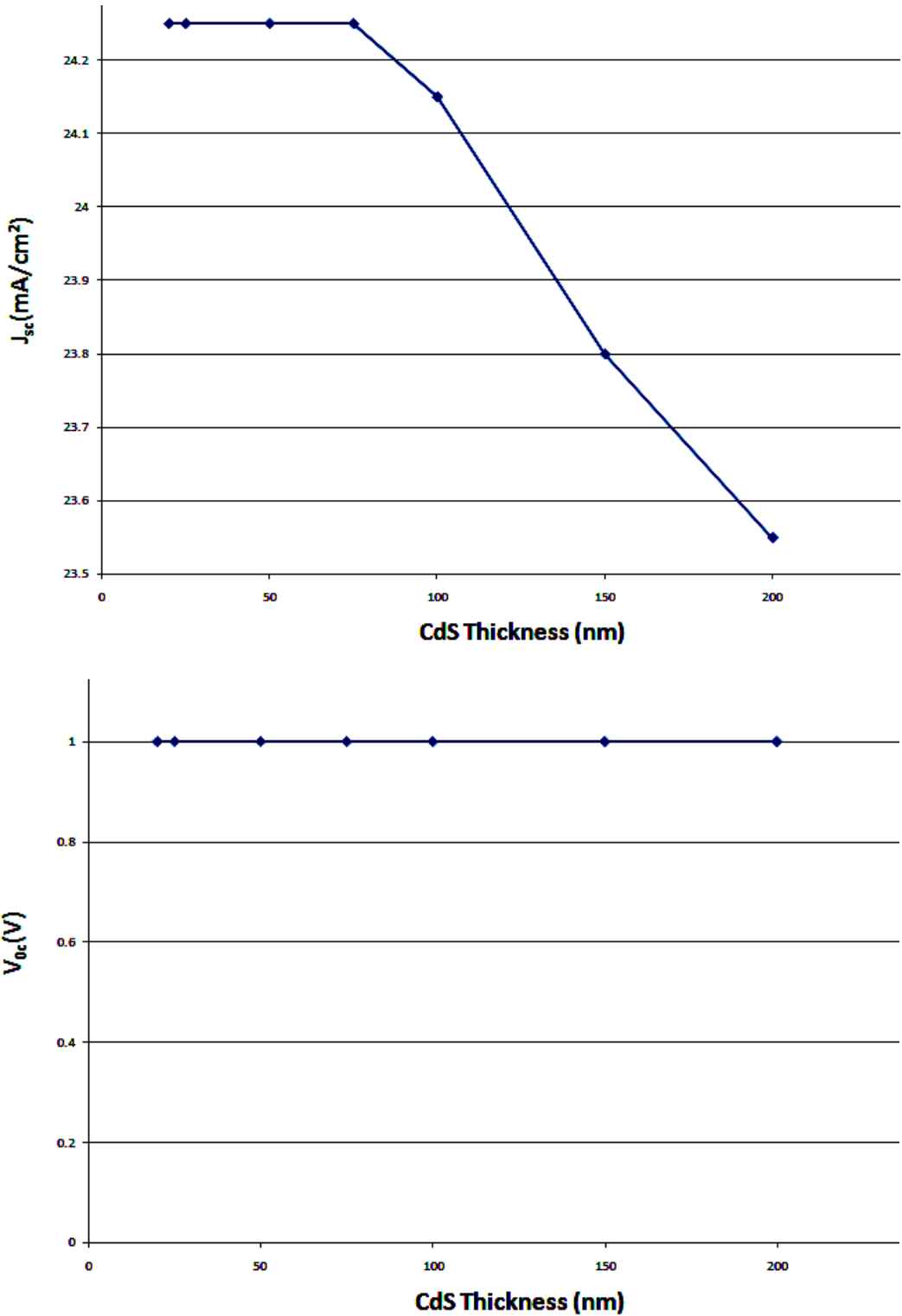

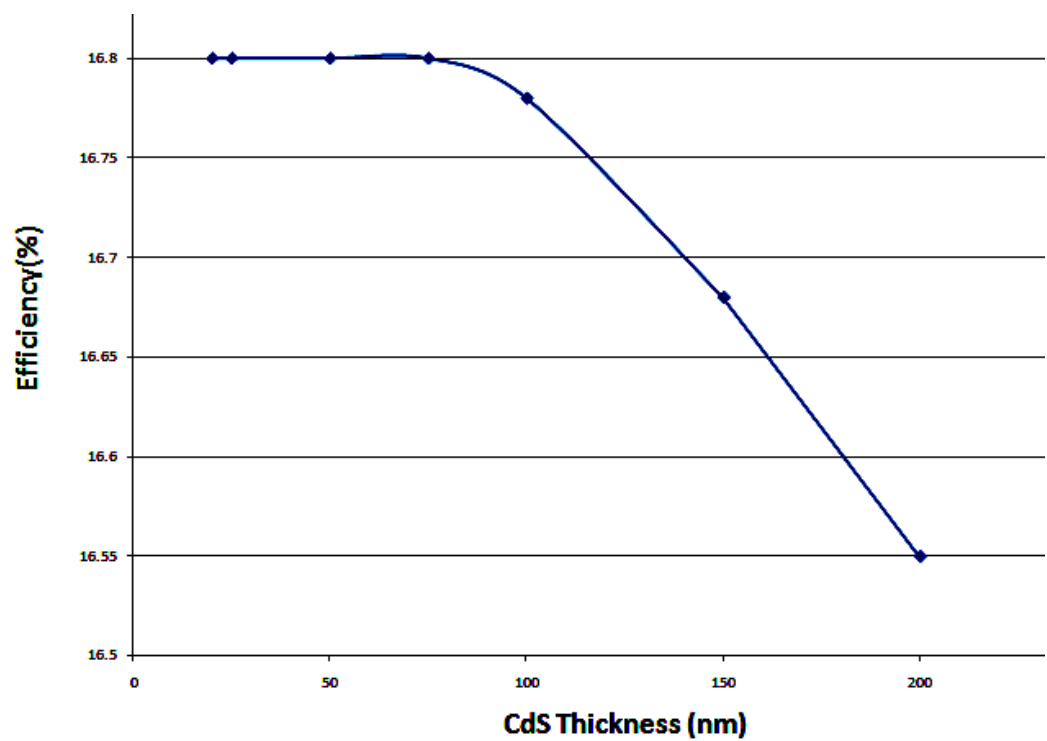

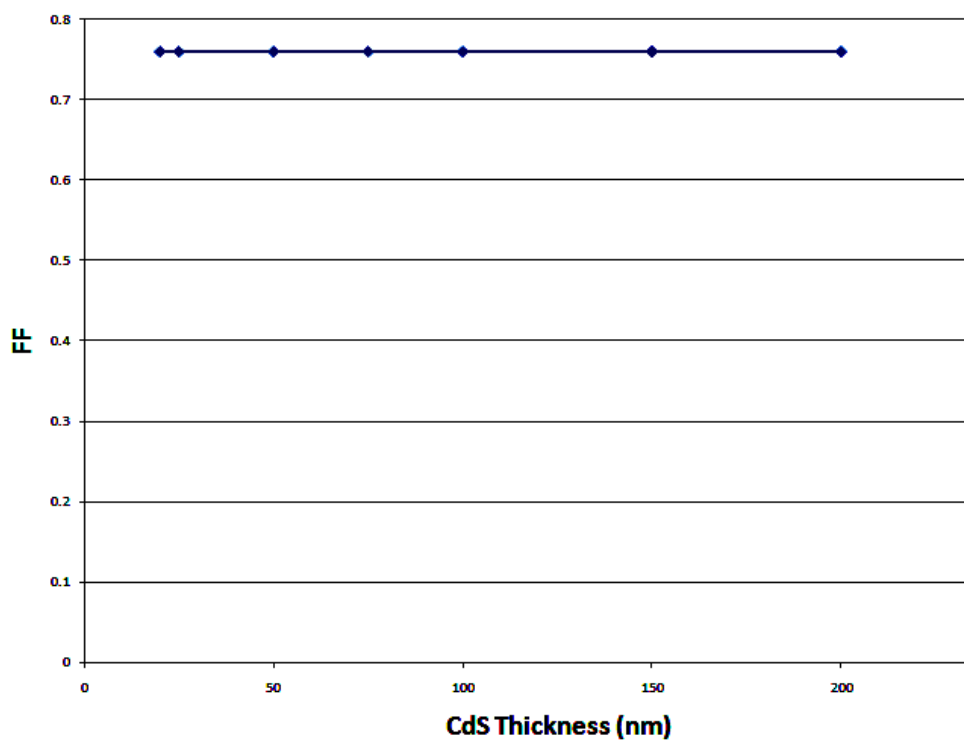

## MULTIJUNCTION TANDEM CdS/CdTe SOLAR CELLS

Multijunction solar cells are created from multiple p-n junctions of different semiconductor materials with different band gap, so that they have the ability to absorb most of the energy of the solar spectrum. Another way to increase the efficiency of solar cells is the use of such elements, each of which uses a certain part of the spectrum of solar radiation for the production of electric current. Tandem solar cells can be used as single or serial connection, where the current in both cases is similar. The structure of the solar cell in the series connection is very simple, but because of limitations caused by wide bandgap, the solar cells of a single connection, from the viewpoint of efficiency, are more optimal. The most common method of manufactiriong of tandem solar cells is their growing even when the layers are built up sequentially on the substrate and provide tunneling contact layers in individual cells. Due to the increasing of number of band gaps, there is an increase in the efficiency of an element. The upper part of the element has the largest band gap, so it absorbs photons, which have higher energy of the spectrum of the incident light, while the lower part of the element has a small bandgap width, and hence, there absorption of the low energy photons takes place [24].

In order to increase the effeciency and to determine the most optimal element will use single-junction solar cells, which we have discussed in previous sections, connected back to back. The parameters of these elements are known to us and optimal performance elements had previously been simulated. Figure 4 shows the proposed order of the layers in this solar cell. This element is constructed of two layers of CdS, one of which is p-type, and the other is n-type, and two layers of CdTe, also of p- and n-type. The outer layer of this element is a layer of ITO of 200 nm thick to provide greater absorption of light flux, and the lower layer is a copper layer with the thickness of 500 nm to reflect light flux.

| ITO |
|---|
| p- CdTe (50нм) |
| n- CdTe (200нм) |
| p- CdS (300нм) |
| n- CdS (1000нм) |
| Cu |

Figure 4. Schematic representation of multi-junction tandem CdS/CdTe solar cells

We perform numerical simulations and determine the most effective solar cells by use of AMPS-1D software, all parameters of simulation associated with each layer listed in Table 2.

After the numerical simulation and obtaining of current-voltage characteristics of the solar cell, the following results

A) On the first stage in order to obtain more effective thickness of the top layer of the CdS/CdTe solar cell, thickness of CdS p-layer was varied from 20 nm to 100 nm. Since this layer is disposed on the top surface under the absorbent layer, hence, it has a minimum thickness as compared with other layers. Thus the optimal thickness of 50 nm for the layer, ensuring the highest efficiency was obtained.

Table 2 .: The parameters of the materials used in the simulation of the tandem multijunction CdS / CdTe solar cell

| Material | Band gap (eV) | Conductivity type | Conduction Band | Valence Band | Electron Affinity (eV) | Electron Mobility (cm2 /v/s) | Hole Mobility (cm2 /v/s) | Free Carrier Concentration (cm$^{-3}$) | Relative Permittivity |
|---|---|---|---|---|---|---|---|---|---|
| ITO | 3.60 | N | 2.0*1020 | 1.8*1019 | 4.10 | 50.0 | 70.0 | 1.0*1020 | 2.0 |
| CdS | 2.40 | P | 2.2*1018 | 1.8*1019 | 4.0 | 25.0 | 100.0 | 1.0*1016 | 10.0 |
| CdS | 2.40 | N | 2.2*1018 | 1.8*1019 | 4.0 | 25.0 | 100.0 | 1.1*1018 | 10.0 |
| CdTe | 1.50 | P | 8.0*1017 | 1.8*1019 | 3.90 | 40.0 | 320.0 | 2.0*1014 | 9.4 |
| CdTe | 1.50 | N | 8.0*1017 | 1.8*1019 | 3.90 | 40.0 | 320.0 | 2.0*1016 | 9.4 |

Figure 5. The dependence of the output parameters (short-circuit current density $J_s$, the open circuit voltage $V_{oc}$, the fill factor FF and efficiency EFF) of multi-junction tandem CdS/CdTe solar cells on the thickness of CdS p-layer

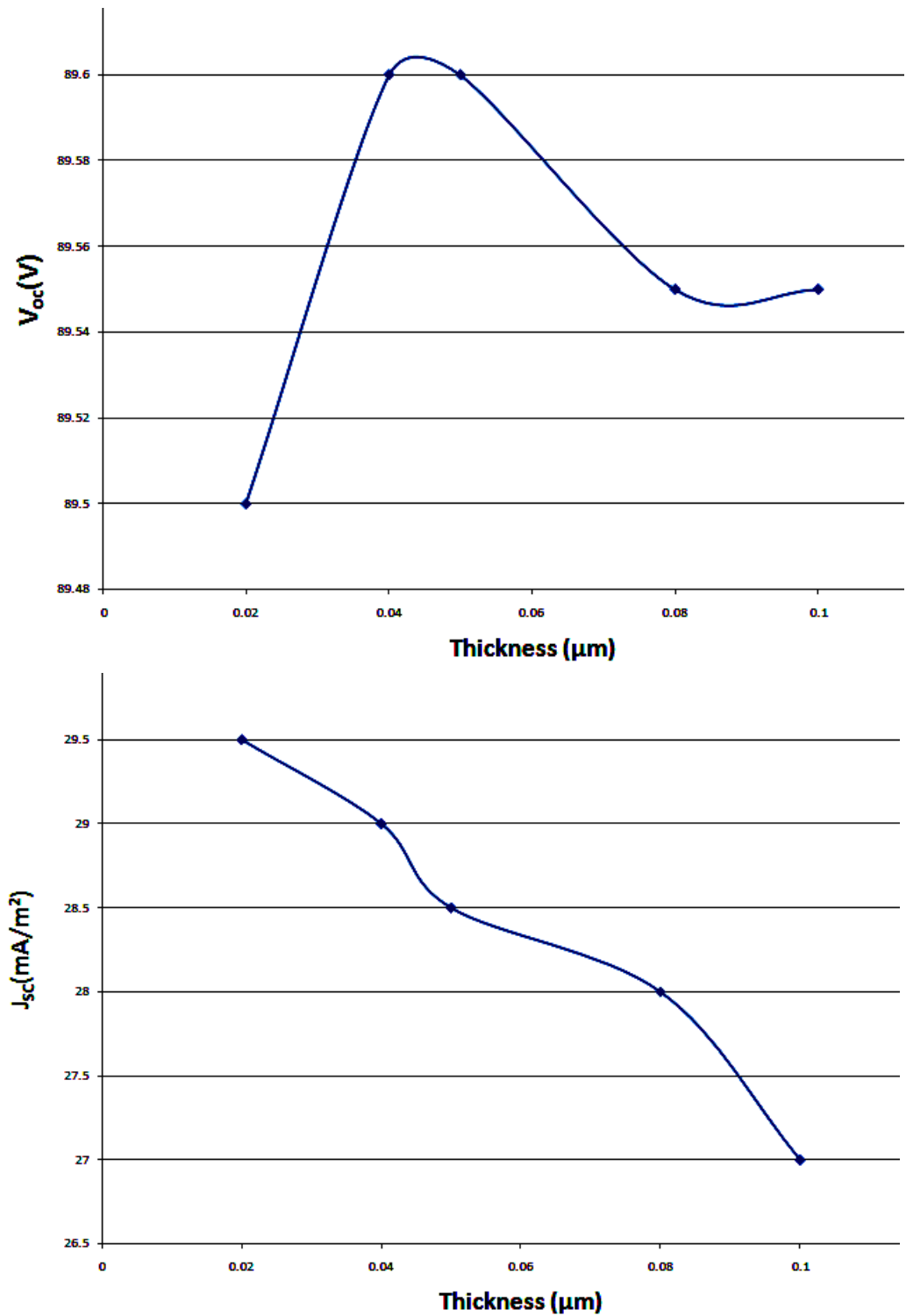

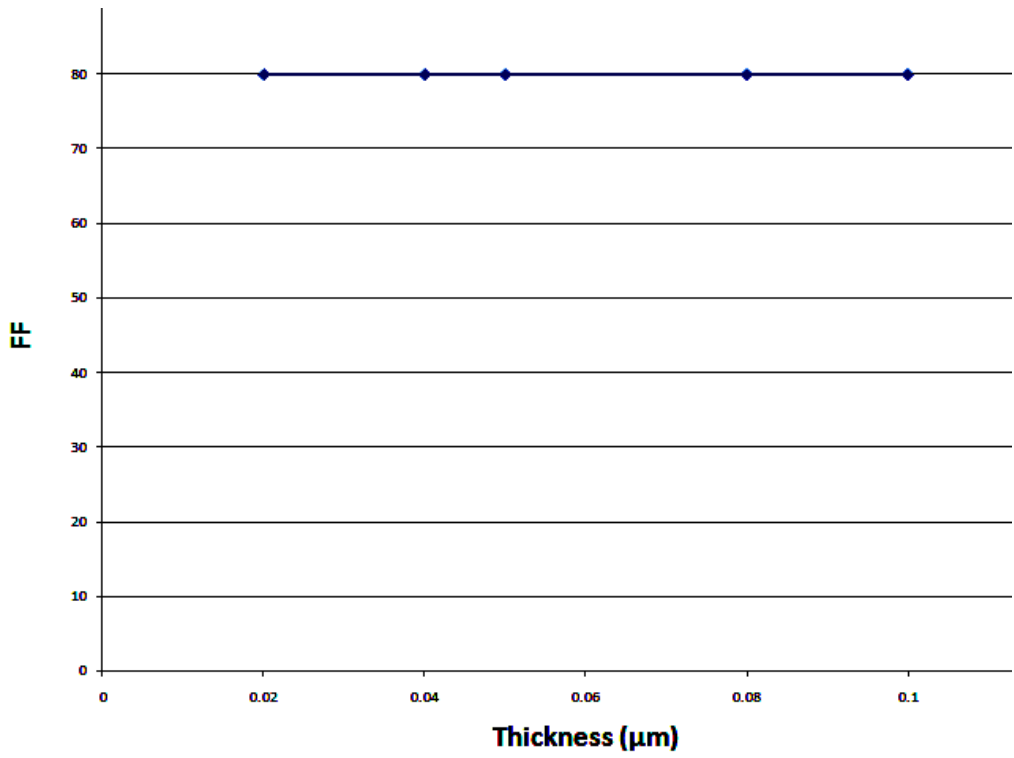

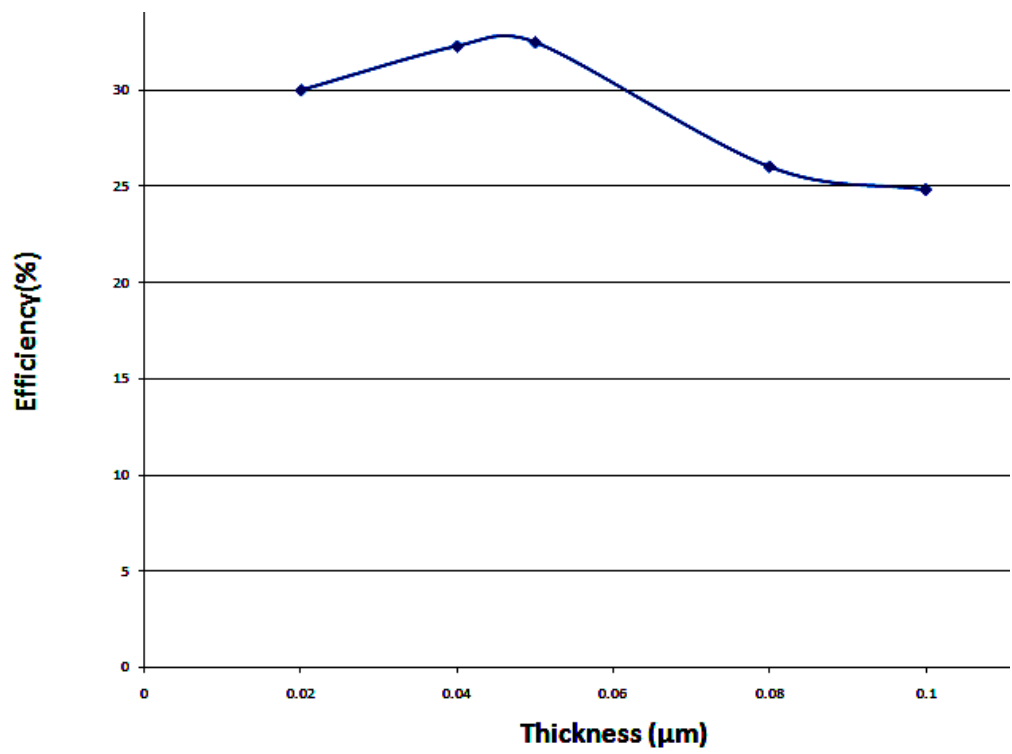

B) At the next stage the thickness of CdS p-layer was kept constant at 50 nm and the thickness of CdS n-layer was varied from 60 nm to 600 nm. Since this layer is disposed on the CdTe p-layer and the CdS p-layer, it should have a thickness less than the thickness of the lower layer, and greater than the thickness of the top layer. Analysis of the results shows that the optimal thickness is 200 nm as shown on the Figure 6.

At both stages of modeling the thickness of the CdTe n-layer was fixed and equal to 3000 nm, and the thickness of the CdTe p-layer remained constant and equal to 1000 nm. The results showed that the optimal efficiency of the tandem multijunction CdS/CdTe solar cell with the above layer thicknesses equal to 31.8%, i.e. efficiency of about two times greater than the same of the single-junction solar cell, under the standard light spectrum of AM 1.5.

Figure 6. The dependence of the output parameters (short-circuit current density Js, the open circuit voltage Voc, the filling factor FF and efficiency EFF) of tandem CdS/CdTe multijunction solar cell on the thickness of CdS n-layer

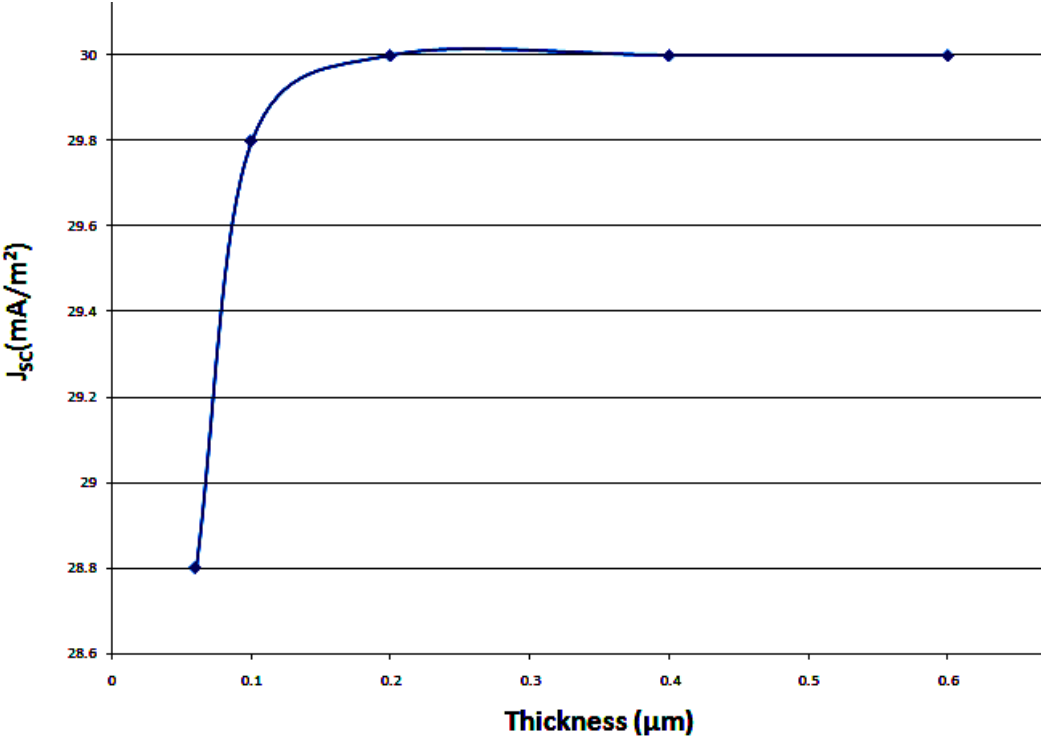

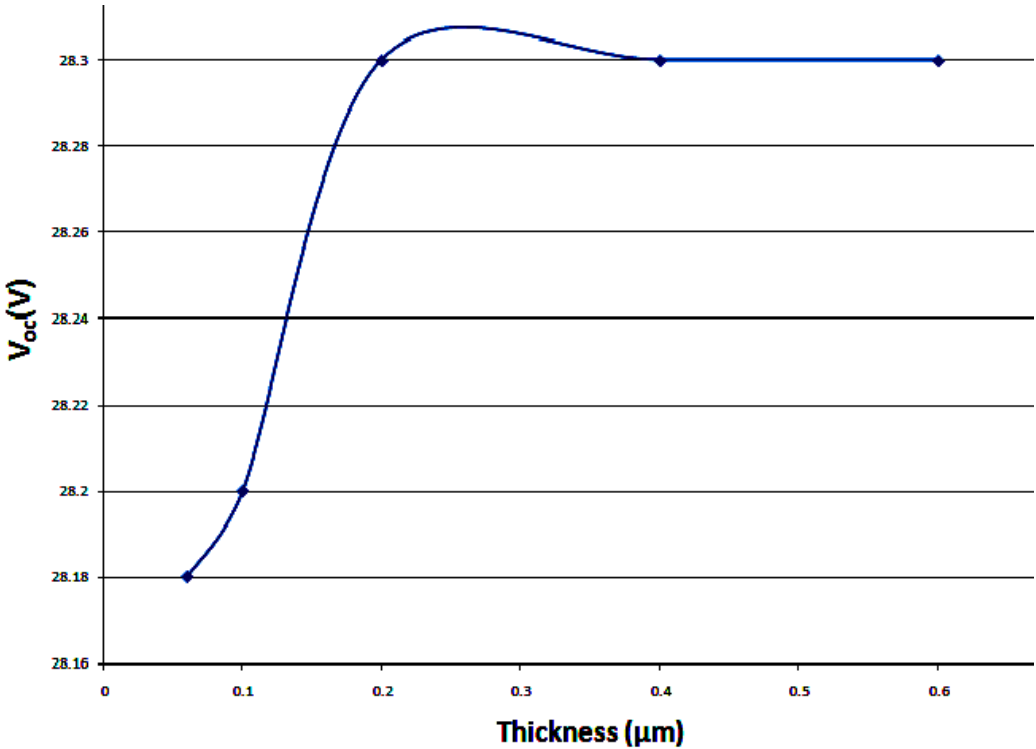

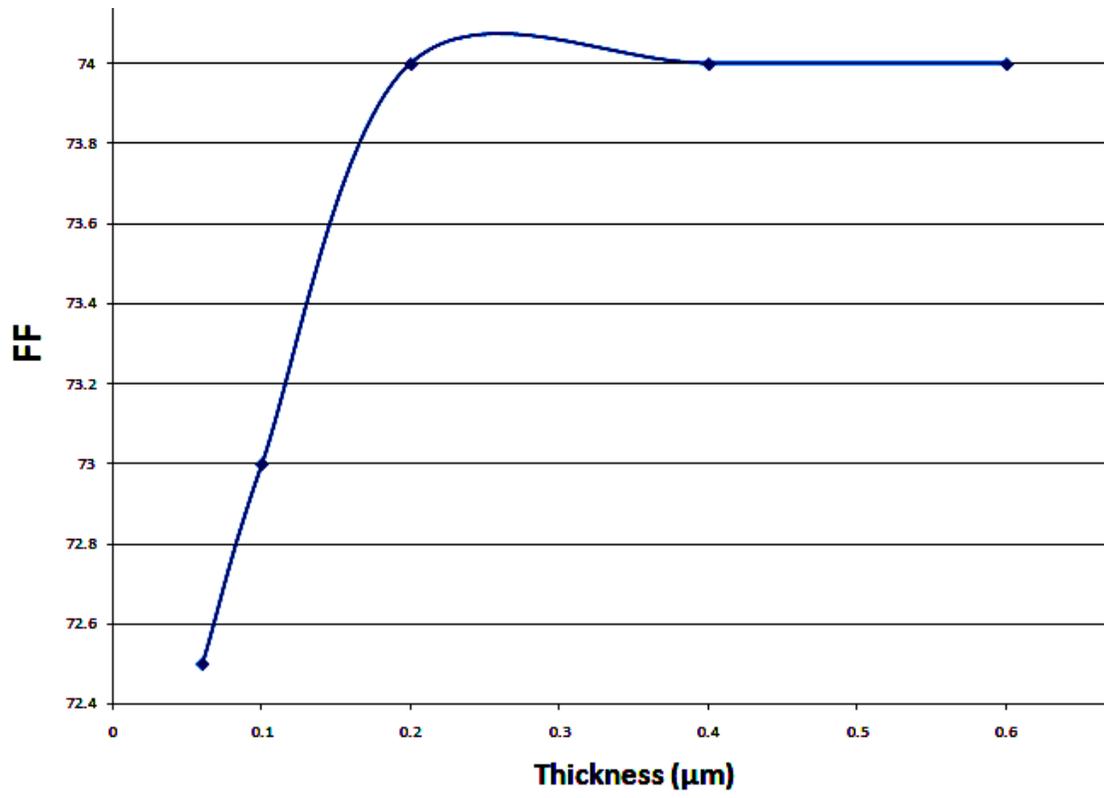

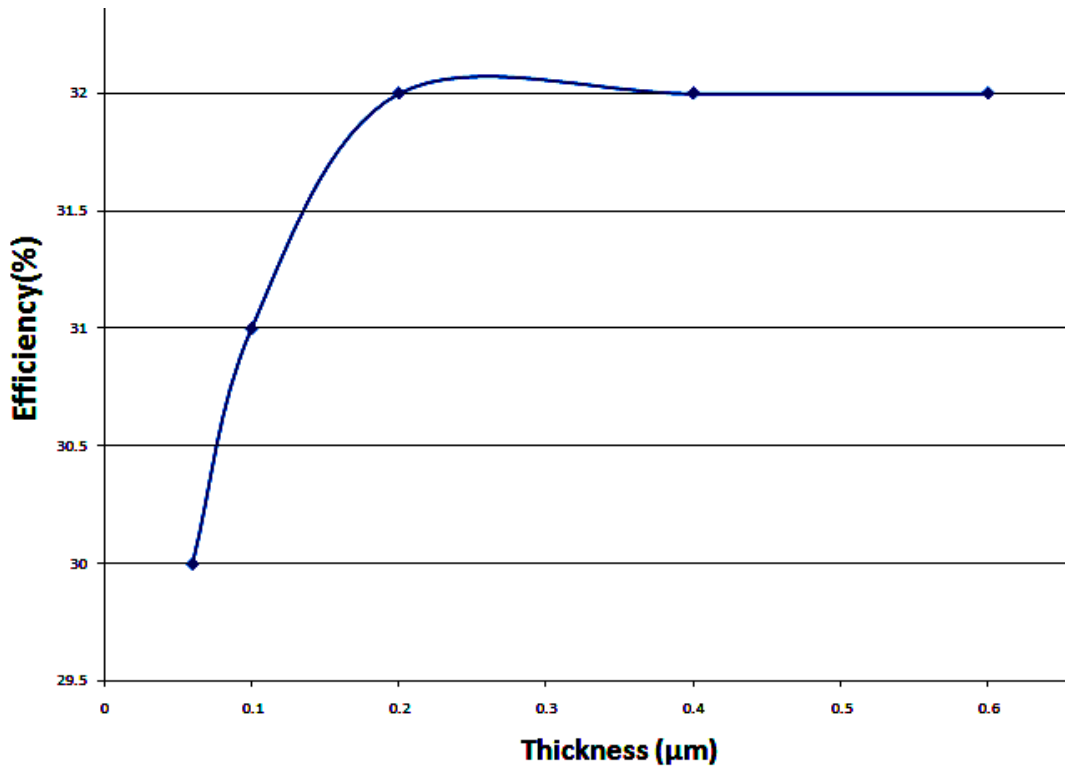

CONCLUSION

Numerical simulation by means AMPS-1D shows that the highest efficiency of single-junction CdS/CdTe solar cell, equal to 18.3%, achieved with a CdTe layer thickness equal to 1000 nm and CdS layer thickness of 60 nm. This result is in good agreement with the experimental data obtained by Chou et al., [17], where the maximum efficiency of CdS/CdTe solar cells was 17.3%.

On the basis of simulation single-junction solar cells we developed an optimal structure of multi-junction tandem CdS/CdTe solar cell, and numerical simulations show that its optimal efficiency in 31.8% can be obtained at the thickness of CdS p-layer equal to 50 nm, and the thickness of the CdS n-layer of 200 nm, while we kept the fixed thickness of the CdTe n-layer of 3000 nm and CdTe p-type layer, equal to 1000 nm.

LITERATURE

[1] Xuanzhi Wu.High-efficiency polycrystalline CdTe thin-film solar cells// Solar Energy, 2004.-V. 77.- P. 803–814.

[2] Nowshad A., Akira Y., Makoto K. Effect of ZnTe and CdZnTe alloys at the Back Contact of 1-μm Thick CdTe Thin Film Solar Cells // Japanese journal of applied physics, 2002.-V.41.-N.5R.- P. 2834-2840.

[3] Sharafat H., Nowshad A., Matin M. A. A numerical study on the prospects of high efficiency ultra thin ZnxCd1-xS/CdTe solar cell// Chalcogenide Letters, 2011.-vol. 8/- n. 3.- P. 263 – 272.

[4] Romeo A., Tiwari A.N., Zogg H. // Proceedings of the 2nd World Conference and Exhibition on Photovoltaic Solar Energy Conversion,1998.- Vienna, Austria.- P. 1105-1108.

[5] Tyan Y.S., Perez-Albuerne E.A.Efficient thin film CdS/CdTe solar cells// Proceedings of 16th IEEE Photovoltaic Specialists Conference, IEEE Publishing, New York, 1982.-P. 794-798.

[6] Ferekides C., Britt J., Ma Y., Killian L. High efficiency CdTe solar cells by closespaced sublimation // Proceedings of 23rd IEEE Photovoltaic Specialists Conference, New York, USA, 1993.- P. 389-393.

[7] Xuanzhi Wu.High-efficiency polycrystalline CdTe thin-film solar cells// Solar Energy, 2004.-N. 77. - P. 803–814.

[8] Romeo N., Bosio A., Romeo A. An innovative process suitable to produce high-efficiency CdTe/CdS thin-film modules // Solar Energy Materials & Solar Cells, 2010.-V. 94.- N. 1.- p. 2–7.

[9] Ferekides C., Britt J., Ma Y., Killian L.High efficiency CdTe solar cells by close spaced sublimation // in Proceedings of the 23rd IEEE Photovoltaic Specialists Conference, 1993.-New York, USA.-P. 389 –393.

[10] Sharafat Hossain.MD, Nowahad. A, Razyokov.T .Prospects Of Back Contacts with Back Surface Fields In High Efficiency $Zn_xCd_{1-x}S$/CdTe solar cells From Numerical Modelling // Chalcogenide Letters,2011.-V. 8.- N. 3.- P. 199 – 210.

[11] Sharafat Hossain. MD, Nowahad A., Matin M.A. A Numerical Study on The Prospects of High efficiency Ultra Thin $Zn_xCd_{1-x}S$/CdTe solar cell //Chalcogenide Letters,2011.- V. 8.- N. 3.- P. 263 – 272.

[12] Wolden C.A., Kurtin J., Baxter J.B., Repins I., Shaheen S.E., Torvik J.T., Rockett A.A., Fthenakis V.M., Aydil E.S. Photovoltaic Manufacturing: Present status, future prospects, and research needs/ J. Vac. Sci. Technol., 2011.-V. 29 .-N.3.- P.1-16.

[13] Gessert T.A., Sheldon P., Li X., Dunlavy D., Niles D., Sasala R., Albright S., Zadler B. Studies of ZnTe Back Contacts to Cds/CdTe Solar Cells// 26th IEEE PV Spec. Conf., 1997.- P.419-422.

[14] Romeo N., Bosio A., Canevari V., Podestà A. Recent progress on CdTe/CdS thin film solar cells// Sol. Energy, 2004.-V. 77.- P. 795–801.

[15] Niemegeers A., Burgelman M. Effects of the Au/CdTe Back Contact on IV and CV Characteristics of Au/CdTe/CdS/TCO Solar Cells // J. Appl. Phys.1997.-V. 81.- N. 6.- P. 2881-2886.

[16] Gessert .T.A, Romero M.J., Dhere R.G., Asher S.E. Analysis of the ZnTe: Cu Contact on CdS/CdTe Solar Cells// NREL Tech. Rep, 2003.-P.520-526.

[17] Chou H.C., Rohatgi A., Jokerst N.M., Thomas E.W., Kamra S. Copper Migration in CdTe Heterojunction Solar Cells// J. Electr. Mater, 1996.-V. 25.-N.7.-P.1093–1098.

[18] Gloeckler M., Fahrenbruch A.L., Sites J.R. Numerical Modeling of CIGS and CdTe Solar Cells: Setting the Baseline // 3rd conf. on PV energy conv, 2003.- P. 491 - 494.

[19] Brownson J.R.S., Georges C., Clement C.L. Synthesis of a δ-SnS Polymorph by Electrodeposition // Chem. Mat. 2006 .-V.18. - P.6397-6402.

[20] Xuanzhi Wu.High-efficiency polycrystalline CdTe thin-film solar cells // Solar Energy, 2004.-V. 77.- P. 803–814.

[21] Nowshad A., Kamaruzzaman S.,Makoto K.Numerical modeling of CdS/CdTe and Cds/CdTe/ZnTe solar cells as a function of CdTe thickness // Solar Energy Materials and Solar Cells, 2007.-V. 91.- N.13.- P. 1202-1208.

[22] Nowshad A., Kamaruzzaman S., Yahya M., Zaharim A. Significance of Absorber Thickness Reduction in CdTe Thin Film Solar Cells for Promising Terrestrial Usage – From the Perspective of Numerical Analysis// Proceedings of the 8[th] WSEAS International Conference on POWER SYSTEMS (PS 2008).- P.299-305.

[23] Nowshad A., Isaka T., Okamoto T., Yamada A., Konagai M. Prospect of Thickness Reduction of the CdTe Layer in Highly Efficient CdTe Solar cells

Towards 1µm // Japanese Journal of Applied Physics,1999.-V. 38 .-N.8R.- P. 4666-4670.

[24] http://www.pveducation.org/pvcdrom/solar-cell-operation/tandem-cells